\documentclass[Physsubmission, Phys]{SciPost}

\hypersetup{
    colorlinks,
    linkcolor={red!50!black},
    citecolor={blue!50!black},
    urlcolor={blue!80!black}
}

\usepackage[bitstream-charter]{mathdesign}
\DeclareSymbolFont{usualmathcal}{OMS}{cmsy}{m}{n}
\DeclareSymbolFontAlphabet{\mathcal}{usualmathcal}

\begin{document}

\begin{center}{\Large \textbf{
Proton reconstruction with the Precision Proton Spectrometer (PPS) in Run 2 and the PPS at HL-LHC \\
}}\end{center}

\begin{center}
Fabrizio Ferro\textsuperscript{1}\\
on behalf of CMS Collaboration
\end{center}

\begin{center}
{\bf 1} INFN, Sezione di Genova (Italy)
\\
* fabrizio.ferro@ge.infn.it
\end{center}

\begin{center}
\today
\end{center}


\definecolor{palegray}{gray}{0.95}
\begin{center}
\colorbox{palegray}{
  \begin{tabular}{rr}
  \begin{minipage}{0.1\textwidth}
    \includegraphics[width=22mm]{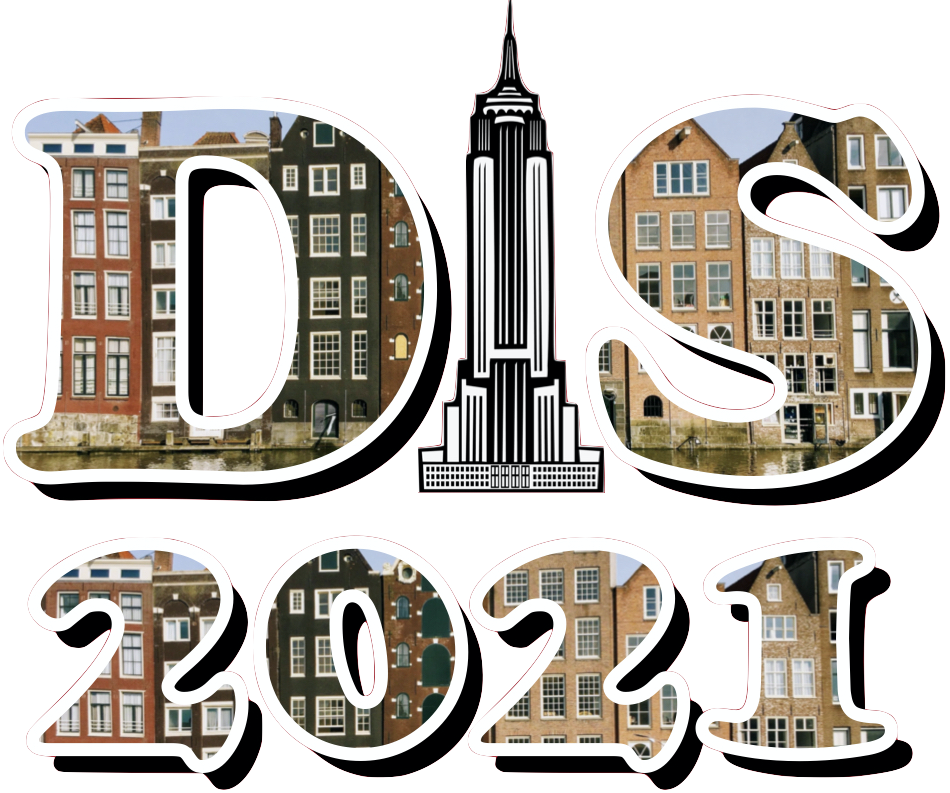}
  \end{minipage}
  &
  \begin{minipage}{0.75\textwidth}
    \begin{center}
    {\it Proceedings for the XXVIII International Workshop\\ on Deep-Inelastic Scattering and
Related Subjects,}\\
    {\it Stony Brook University, New York, USA, 12-16 April 2021} \\
    \doi{10.21468/SciPostPhysProc.?}\\
    \end{center}
  \end{minipage}
\end{tabular}
}
\end{center}

\section*{Abstract}
{\bf
The Precision Proton Spectrometer (PPS) started operating in 2016 and has collected more than 110 fb$^{-1}$ of data over the course of the LHC Run 2, now fully available for physics analysis. This contribution covers the key features of the PPS alignment and optics calibration, which have been developed from scratch. The reconstructed proton distributions, the performance of the PPS simulation and finally the validation of the full reconstruction chain with physics data (dilepton events) are also illustrated.
}

\vspace{10pt}
\noindent\rule{\textwidth}{1pt}
\tableofcontents\thispagestyle{fancy}
\noindent\rule{\textwidth}{1pt}
\vspace{10pt}

\section{Introduction}
\label{sec:intro}
The Precision Proton Spectrometer (PPS) \cite{PPS} detector system has been installed and integrated into
the CMS \cite{CMS} experiment during the LHC Run 2. It is a joint project of the CMS and TOTEM \cite{TOTEM} collaborations with the capability to measure protons scattered at very small angles, and to operate at high instantaneous luminosity. The scattered protons remain inside the beam pipe, displaced from the central beam orbit, and can be measured by detectors placed inside movable beam pipe insertions, called Roman Pots (RP), which approach the beam down to a few mm. The PPS detectors have collected data corresponding to an integrated luminosity larger than 110 fb$^{-1}$ during the LHC Run 2, between 2016 and 2018.
\par The physics motivation behind the PPS detector is the study of central exclusive production (CEP), i.e. the process $pp\rightarrow p^{(*)} + X + p^{(*)}$ mediated by color-singlet exchanges (photons, Pomerons, Z bosons), with at least one of the outgoing protons detected in PPS.
\par
\begin{figure}[h]
\centering
\includegraphics[width=0.95\textwidth]{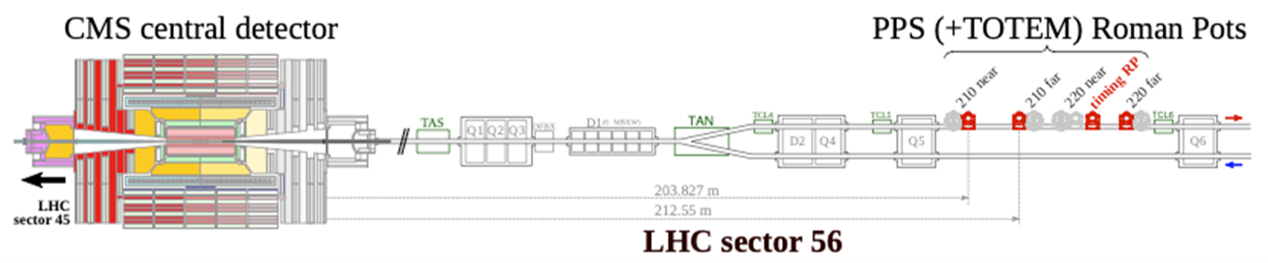}
\caption{Experimental setup of PPS in CMS.}
\label{PPSsetup}
\end{figure}

 Figure \ref{PPSsetup} shows the layout of the RP system installed at around 200–220 m from the CMS interaction point (IP5), along the beam line in LHC sector 56. A symmetric set of detectors is installed in LHC sector 45 at the opposite side of CMS. The RPs approach the beam vertically from the top and bottom, as well as horizontally. During standard machine operation, scattered protons undergo a large displacement in the horizontal direction and a small vertical displacement at the RP positions. The horizontal RPs are hence used. The vertical RPs are used in special configurations of the machine and in low luminosity proton-proton fills for the calibration and alignment of the detectors.

\par Each detector arm consists of two RP stations instrumented with silicon tracking detectors that measure the transverse displacement of protons with respect to the beam, and one RP station with timing detectors to measure their time-of-flight. Silicon strip sensors with a reduced insensitive region on the edge facing the beam were initially used (10 planes per pot), but since they cannot sustain a large radiation dose and cannot identify multiple tracks in the same event, they have been gradually replaced by new 3D silicon pixel sensors (6 planes per pot).
\par The proton reconstruction relies on the alignment of the detector planes with respect to the LHC beam and on the knowledge of the transport matrices parametrizing the LHC magnet
lattice. The latter, referred to as the beam optics, relate the transverse position and direction of a proton track along the beam line to the transverse position of the proton at the nominal interaction point ($x^*, y^*$), the proton scattering angles in the horizontal and vertical directions ($\theta_x^*,\theta_y^*$) and its momentum p, or rather its fractional momentum loss $\xi = (p_{nom}-p)/p_{nom}$, where $p_{nom}$ is the nominal beam momentum.

\section{Track reconstruction}

The Data Acquisition System of the detectors provides as input of the reconstruction chain the coordinates of the pixels that have collected a charge above a certain threshold, set by a calibration procedure, and a measurement of the collected charge in terms of ADC counts.

The set of pixel fired in the event undergoes a clusterization procedure that collects in sub-sets, called clusters, the adjacent fired pixels. From any cluster a single local reconstructed hit is generated, whose position is evaluated by means of a charge sharing algorithm. 

The track reconstruction is carried out on a single pot basis, using the set of hits provided by the six planes. The procedure is iterative: it starts by fitting all the hit combinations from the maximum number of planes available in a region of interest (usually 6) and choosing the one that provides the minimum $\chi^2$ over degree of freedom. If a track candidate with a $\chi^2$ under a certain threshold is found, the track is saved, the used hits removed and the iterative procedure restarted.  

\section{Alignment}
The alignment procedure involves multiple steps. A special alignment calibration LHC fill
allows to determine the absolute position of the RPs with respect to the beam. This
calibration then serves as a reference for the alignment of every physics fill with standard
conditions. Once the tracking RPs are aligned with respect the beam, the timing
RPs are aligned with respect to the tracking RPs.
The alignment procedures depend on the LHC settings (optics, collimators etc.). Therefore separate analyses are performed for periods with different settings. The settings are often changed during technical stops (TS), which are also used to adjust the RP configuration, possibly influencing the alignment results.
\par The alignment fill is a calibration fill with standard optics but low beam intensity which
serves multiple purposes. Primarily, it allows to establish the RP position with respect to
the LHC collimators in a procedure analogous to the LHC collimator alignment. 
This is a precondition for systematic RP insertion close to the high-intensity LHC
beams. Then, thanks to the low intensity, one can insert both horizontal and vertical RPs very
close to the beam, at about 5$\sigma_{beam}$, where $\sigma_{beam}$ is the transverse size of the beam.  This yields an overlapping system of detectors 
that allows for relative alignment among the RPs in each arm.
The relative alignment, among sensor planes in all RPs and among all RPs in one arm, is
determined by minimising residuals between hits and fitted tracks.


\par For each high-luminosity LHC fill (physics fill), the horizontal alignment is obtained by
matching observations from the fill to those from the reference alignment fill. The chosen matching metrics is the slope of the profile of $(y_F-y_N)$ as a function of $x$,
where $y_N$ and $y_F$ stand for the vertical track positions in the near and far RP with respect to the IP. The vertical alignment is obtained by extrapolating the observed vertical profile of $x_N - x_F \textrm{ vs. } x_N$ to the horizontal beam position.

\section{Optics calibration}
Scattered protons are detected in the PPS RPs after having traversed a segment of the
LHC lattice containing 29 main and corrector magnets per beam. The transverse proton position at the RP location with
transverse position ($x^*, y^*$) and angles ($\theta_x^*, \theta_y^*$) at the IP is described at first approximation as 
$$ x = v_x(\xi)\cdot x^* + L_x(\xi) \cdot \theta_x^* + D_x(\xi)\cdot \xi$$ 
$$ y = v_y(\xi)\cdot y^* + L_y(\xi) \cdot \theta_y^* + D_y(\xi)\cdot \xi.$$ 
Therefore the reconstruction of the proton kinematics at the IP from the measurements at the RPs needs a precise knowledge and calibration of the LHC optics.
\par The first step of the optics calibration is to update the nominal transport model using LHC
databases. The proper version of the LHC magnet lattice description for the data taking period under consideration is used, and the nominal magnet strength file for a given beam optics is always updated using measured data. 
For the evaluation of the proton kinematics, the horizontal dispersion $D_x$ is the most important, because
it allows to convert the x-coordinate of the proton impact point into the fractional proton momentum loss $\xi$. The determination of $D_x$, using proton trajectories measured in the RPs, is described in details in \cite{optics}. The 2017 and 2018 optics calibration procedure goes a step further and exploits (semi)-exclusive $\mu\mu$ production in which the exclusivity of the process plays a key role.
\par In a third step the vertical dispersion $D_y$ is determined from minimum bias RP data. The
calibration of the dispersion functions is followed by the calibration of the remaining optical
functions in the transport equations, namely the horizontal and vertical effective
lengths $L_{x,y}(\xi)$, and the corresponding magnification functions $v_{x,y}$; other optical functions are less relevant from the point of view of the proton reconstruction.

\section {Proton reconstruction}
The proton reconstruction consists of back-propagating the protons from the RPs, where they
are measured, to the IP, where the kinematics is determined. The propagation follows the LHC optics; inputs to it are the proton tracks detected by the RPs and aligned with respect to the LHC beam. Two complementary reconstruction strategies are used: single-RP and multi-RP.
\par The single-RP reconstruction is a simple approach that uses information from single RPs only.
Because of the reduced input information, only $x$ and $\theta_y^*$ can be estimated via
$$ \xi \sim x/D_x, \hspace{1cm}\theta_y^* = y/L_y(\xi) .$$
These equations reflect only the leading terms from the optics decomposition, which implies a degraded resolution. On the other hand, notable advantages of this approach are the applicability even when the proton
track is not available in the other RP(s) of the arm and a slightly smaller systematic uncertainties with respect to the multi-RP method.
\begin{figure}[h]
\centering
\includegraphics[width=0.8\textwidth]{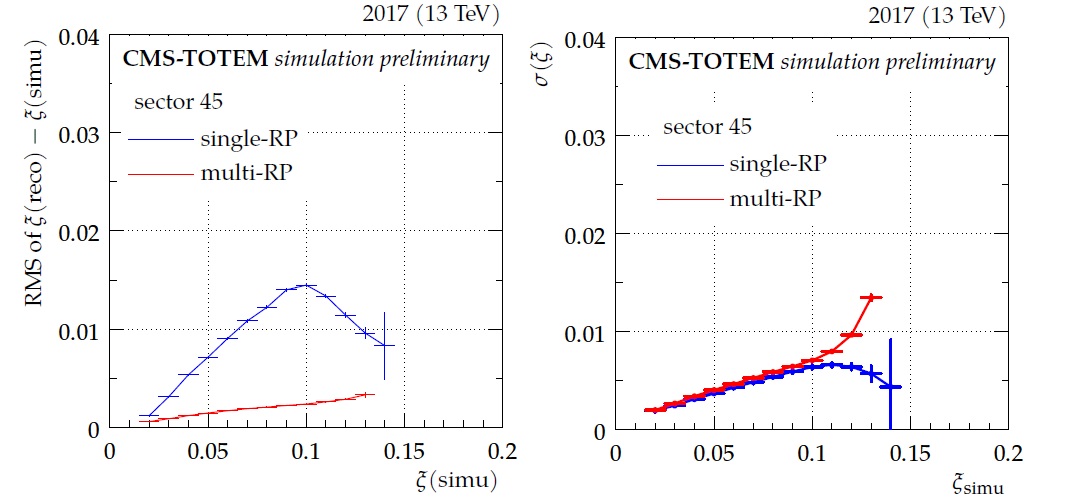}
\caption{Comparison between single and multi-RP reconstruction using simulated data: resolution (left) and systematic uncertainties (right). }
\label{reco}
\end{figure}
\par The multi-RP reconstruction aims at exploiting the full potential of the spectrometer: it searches
for proton kinematics that matches best the observations from all RPs and all projections by
minimising the following function:
$$\chi^2 = \sum_{i:RPs} \sum_{q:x,y}\left( \frac{d^i_q-(T^iq^*)_q}{\sigma_q^i}\right)^2$$
where the vector $d^i$ is the proton position at the i-th RP, $\sigma_q^i$ its uncertainties, the vector $d^*$ denotes the proton kinematics at the IP and the matrix
$T^i$ is for the proton transport between the IP and the i-th RP. The performance differences between the two methods are summarized in Figure \ref{reco} where simulation studies are reported.

\section{Future of PPS}
PPS is preparing the installation of new detectors for LHC Run 3 that is scheduled to start at the beginning of 2022. With respect to the setup used in Run 2 an additional RP for timing measurements has been installed on both sides. Moreover, another remarkable change is the installation of motors that will allow the vertical movement of the pixel detectors, independently of the RP position, in order to change the detector part facing the beam (that undergoes most of the radiation damage), extending therefore the overall duration of the sensor. 
\par An {\it Expression of Interest} for a proton spectrometer to be built in the High Luminosity phase of LHC (HL-LHC) has been published \cite{eoi}. It would extend the sensitivity of the spectrometer to CEP to lower masses and to lower production cross-sections both in Standard Model measurements and in searches for new physics.    

\section{Conclusion}
The PPS detector of CMS has taken data during Run 2 and has provided the kinematic parameters of the tagged proton by means of the reconstruction strategy shortly described in this article. The experience achieved so far will be exploited in Run 3 where PPS will take data with an updated version of its detectors. An outlook on the longer term future of proton spectroscopy at LHC has been given as well.


\bibliography{SciPost_Example_BiBTeX_File.bib}

\nolinenumbers

\end{document}